\documentclass[aps,prl,twocolumn,showpacs,superscriptaddress,groupedaddress]{revtex4-1}  
\usepackage{graphicx, color}  
\usepackage{dcolumn}   
\usepackage{bm}        
\usepackage{amssymb}   
\usepackage{amsmath}   
\usepackage{multirow}
\usepackage{enumerate}
\usepackage{comment}
\usepackage{tabularx}
\usepackage{booktabs}
\usepackage{placeins}
\usepackage{xcolor}
\usepackage{hyperref}

\begin{document}



\title{Frequency-Dependent Squeezed Vacuum Source for Broadband Quantum Noise Reduction in Advanced Gravitational-Wave Detectors}

\author{Yuhang Zhao$^{1,2}$} 
\author{Naoki Aritomi$^{3}$} 
\author{Eleonora Capocasa$^{1}$} 
\email[Corresponding author: ]{eleonora.capocasa@nao.ac.jp}
\author{Matteo Leonardi$^{1}$} 
\email[Corresponding author: ]{matteo.leonardi@nao.ac.jp}
\author{Marc Eisenmann$^{4}$} 
\author{Yuefan Guo$^{5}$}
\author{Eleonora Polini$^{4}$} 
\author{Akihiro Tomura$^{6}$} 
\author{Koji Arai$^{7}$} 
\author{Yoichi Aso$^{1}$} 
\author{Yao-Chin Huang$^{8}$} 
\author{Ray-Kuang Lee$^{8}$} 
\author{Harald L{\"u}ck$^{9}$} 
\author{Osamu Miyakawa$^{10}$} 
\author{Pierre Prat$^{11}$} 
\author{Ayaka Shoda$^{1}$} 
\author{Matteo Tacca$^{5}$} 
\author{Ryutaro Takahashi$^{1}$} 
\author{Henning Vahlbruch$^{9}$} 
\author{Marco Vardaro$^{5,12,13}$} 
\author{Chien-Ming Wu$^{8}$} 
\author{Matteo Barsuglia$^{11}$} 
\author{Raffaele Flaminio$^{4,1}$}

\address{$^{1}$National Astronomical Observatory of Japan, 2-21-1 Osawa, Mitaka, Tokyo, 181-8588, Japan}
\address{$^{2}$The Graduate University for Advanced Studies(SOKENDAI), 2-21-1, Osawa, Mitaka, Tokyo 181-8588, Japan}
\address{$^{3}$ Department of Physics, University of Tokyo, 7-3-1 Hongo, Tokyo, 113-0033, Japan}
\address{$^{4}$Laboratoire d'Annecy-le-Vieux de Physique des Particules (LAPP), Université Savoie Mont Blanc, CNRS/IN2P3, F-74941 Annecy-le-Vieux, France}
\address{$^{5}$Nikhef, Science Park, 1098 XG Amsterdam, Netherlands}
\address{$^{6}$The University of Electro-Communications  1-5-1 Chofugaoka, Chofu, Tokyo 182-8585, Japan}
\address{$^{7}$LIGO, California Institute of Technology, Pasadena, California 91125, USA}
\address{$^{8}$Institute of Photonics Technologies, National Tsing-Hua University, Hsinchu 300, Taiwan}
\address{$^{9}$Institut f{\"u}r Gravitationsphysik, Leibniz Universit{\"a}t Hannover and Max-Planck-Institut f{\"u}r Gravitationsphysik (Albert-Einstein-Institut), Callinstra{\ss}e 38, 30167 Hannover, Germany}
\address{$^{10}$Institute for Cosmic Ray Research (ICRR), KAGRA Observatory, The University of Tokyo, Kamioka-cho, Hida City, Gifu 506-1205, Japan}
\address{$^{11}$Université de Paris, CNRS, Astroparticule et Cosmologie, F-75013 Paris, France}
\address{$^{12}$Institute for High-Energy Physics, University of Amsterdam, Science Park 904, 1098 XH Amsterdam, Netherlands}
\address{$^{13}$Università di Padova, Dipartimento di Fisica e Astronomia, I-35131 Padova, Italy}


\begin{abstract} 

The astrophysical reach of current and future ground-based gravitational-wave detectors is mostly limited by quantum noise, induced by vacuum fluctuations entering the detector output port. The replacement of this ordinary vacuum field with a squeezed vacuum field has proven to be an effective strategy to mitigate such quantum noise and it is currently used in advanced detectors. However, current squeezing cannot improve the noise across the whole spectrum because of the Heisenberg uncertainty principle: when shot noise at high frequencies is reduced, radiation pressure at low frequencies is increased. A broadband quantum noise reduction is possible by using a more complex squeezing source, obtained by reflecting the squeezed vacuum off a Fabry-Perot cavity, known as filter cavity. Here we report the first demonstration of a frequency-dependent squeezed vacuum source able to reduce quantum noise of advanced gravitational-wave detectors in their whole observation bandwidth. The experiment uses a suspended 300-m-long filter cavity, similar to the one planned for KAGRA, Advanced Virgo, and Advanced LIGO, and capable of inducing a rotation of the squeezing ellipse below 100 Hz.

\end{abstract}

\maketitle

\section{\label{sec:introduction}Introduction}

Gravitational-wave astronomy, started in 2015 with the first detection of a black hole merger by LIGO \cite{PhysRevLett.116.061102}, has rapidly become a mature research field. After the groundbreaking observation of the binary neutron star coalescence GW170817 \cite{BNS} and the publication of the first gravitational-wave sources catalog \cite{catalog},  the network of advanced interferometric detectors (composed of the two Advanced LIGO \cite{0264-9381-32-7-074001} and Advanced Virgo \cite{0264-9381-32-2-024001}) is now performing a third period of data taking (O3), detecting several candidates per month and sending open alerts to the astronomical community \cite{graceDB}. The Japanese detector KAGRA \cite {PhysRevD.88.043007} is about to join the network and is expected to be followed by a fifth detector, LIGO India, in 2024 \cite{Prospect}. 

One of the main limitations to the detector sensitivity is quantum noise, which originates from the quantum nature of light. It manifests itself in two components:  quantum shot noise, dominating at high frequencies of the detector spectrum and quantum radiation pressure noise, dominating at low frequencies.
In 1981, Caves clarified that quantum noise is originated by quantum vacuum fluctuations entering the interferometer antisymmetric port \cite{PhysRevD.23.1693}. He also suggested that quantum noise could be reduced by replacing ordinary vacuum states with the so-called squeezed vacuum states, whose noise is reshaped in order to reduce it in one quadrature and increase it in the orthogonal one, in accordance with Heisenberg's uncertainty principle. A squeezed state can be represented in the quadrature plane as an ellipse, characterized by two frequency-dependent parameters: the squeezing magnitude (the ratio of the ellipse axes with respect to the unsqueezed state) and the squeezing angle (the ellipse orientation).
 
More than 30 years of experimental developments were needed before such technology could be integrated in the detectors \cite{HG,BHS}. Since the beginning of O3, in April 2019, both Advanced Virgo and Advanced LIGO feature a frequency-independent squeezing source which is enhancing their performances \cite{SQZ_virgo}, bringing an increase in the detection rate of up to $50\%$ \cite{SQZ_LIGO}. 
However, the squeezed vacuum currently used in advanced detectors cannot reduce simultaneously both quantum noise components. The increase of radiation pressure noise due to frequency-independent squeezing has been recently observed in both Advanced Virgo and Advanced LIGO \cite{RP_Virgo, RP_LIGO}. The reason is the optomechanical coupling with the suspended test masses, which induces a rotation of the squeeze ellipse by $90 ^{\circ}$ across the frequency spectrum, so that the noise-reduced quadrature cannot be aligned with the gravitational-wave signal at all frequencies \cite{0295-5075-13-4-003}. Even if the current degradation of the sensitivity induced at low frequency is barely observable due to the presence of technical noises and to the limited amount of power used, it is expected to become more and more detrimental as the detectors approach their design sensitivity.

A possible solution to reduce quantum noise in the whole spectrum is the injection of frequency-dependent squeezed vacuum, where the squeeze ellipse rotates as a function of frequency, to counteract the rotation induced by the interferometer. Such reduction is optimal if the rotation is at the crossover frequency between the radiation-pressure dominated and the shot-noise dominated regions, around 30-70 Hz for advanced gravitational-wave detectors.

 The interaction of a frequency-independent squeezed state with an optical Fabry-Perot resonator, referred to as filter cavity, is able to induce such rotation of the squeeze ellipse \cite{KKK}. In the past, squeeze ellipse rotation has been realized in the megahertz \cite {PhysRevA.71.013806} and kilohertz \cite {PhysRevLett.116.041102} regions using rigid, meter-scale filter cavities.

In this Letter, we report the first demonstration of a frequency-dependent squeezed vacuum source able to reduce quantum noise of advanced gravitational-wave detectors in their whole observation bandwidth. The experiment uses a filter cavity capable of rotating the squeezing ellipse below 100 Hz, in the region needed for advanced detectors. The validation of this technology using a 300-m filter cavity is especially significant as KAGRA, Advanced LIGO, and Advanced Virgo all plan to integrate filter cavities of comparable length in the upcoming upgrading phase \cite{A+,AdVirgo+}.

Concurrent with, and independent of, this work, a similar experiment, carried out in the U.S. with a 16-m filter cavity, shows a 30-Hz rotation frequency and includes a new control system, capable of meeting the stringent noise requirements of LIGO and other gravitational-wave detectors \cite {MITpaper}. 

\section{\label{sec:setup} Experimental setup}     

The experiment is realized at the National Astronomical Observatory of Japan (NAOJ), using the infrastructure of the former TAMA 300 interferometer \cite {TAMA}.

\begin{figure}[h!]
	\centering{\includegraphics[width=0.5\textwidth]{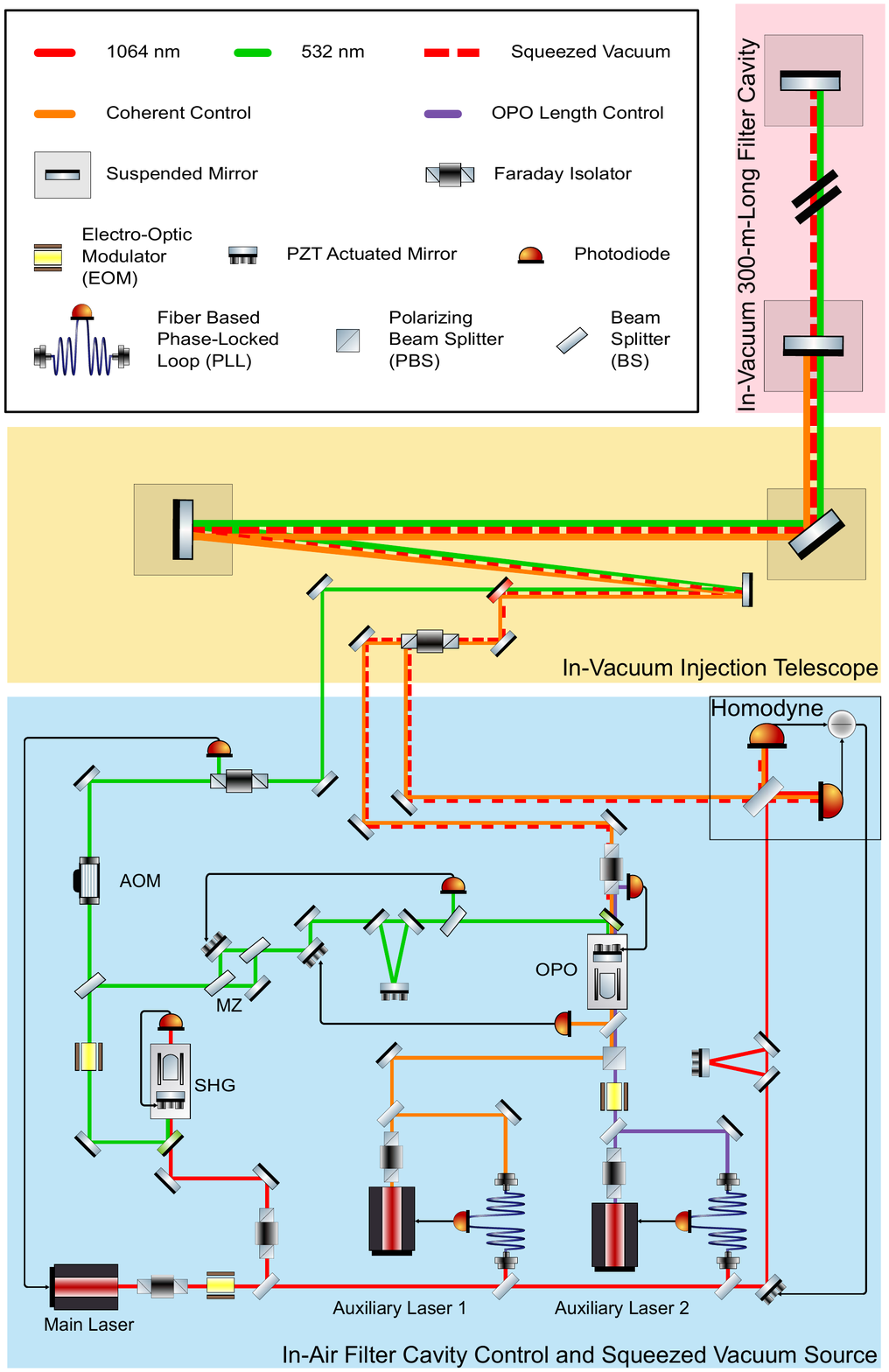}}
	\caption{Schematic diagram of the experimental setup. The squeezed vacuum beam is produced by the OPO and injected into the filter cavity through the in-vacuum injection telescope. The reflected frequency-dependent squeezing is rejected by the in-vacuum Faraday and characterized by the homodyne detector. The green beam produced by the SHG is used to pump the OPO and as auxiliary beam for the control of the filter cavity.}
	\label{tamascheme}
\end{figure}

The experimental setup, sketched in Fig. \ref{tamascheme}, consists of two parts: a source of frequency-independent squeezed vacuum and a suspended filter cavity. The two are connected by an in-vacuum injection system, including suspended mirrors, which couples the squeezing into the filter cavity. A detailed description of the cavity and the injection apparatus can be found in Ref. \cite{capocasa2}.

\subsection{\label{subsec: FISVS} Frequency-independent squeezed vacuum source} 

The in-air squeezed vacuum source, assembled specifically for this experiment, follows the design of the GEO600 squeezer \cite{SQZ_GEO}.  The core part is the optical parametric oscillator (OPO): a linear hemilithic cavity hosting a Periodically Poled Potassium Titanyl Phosphate (PPKTP) crystal in which the squeezed vacuum is produced through a parametric down-conversion process. This requires a pump beam at twice the squeezing frequency, which is produced by injecting a 1064-nm laser into a second harmonic generator cavity (SHG). The main laser, a 2-W 1064-nm Nd:YAG laser, is used to pump the SHG and produce green light (with a wavelength of 532 nm) and as a local oscillator for the balanced homodyne detector \cite{HM}, used to characterize the squeezing. A mode cleaner cavity and a Mach-Zehnder interferometer (MZ) are installed respectively to spatially clean and to stabilize in power the green pump beam before it enters the OPO. Two auxiliary lasers, frequency offset locked with the main laser, are also used. The first one (AUX1), injected into the OPO with a different polarization with respect to the produced squeezed beam, is used to control the OPO length, which is only resonant for 1064-nm wavelength. The second one (AUX2) is also injected into the OPO and copropagates with the squeezed vacuum beam up to the homodyne detector, to track the squeezing phase and lock it with respect to the local oscillator \cite{coco}. Custom analog electronics have been developed to operate the control loops involved in the experiment. An automatic hierarchical control acquisition procedure is also implemented to bring the squeezing source in the working state within a few seconds.
A maximum squeezing level of  $ 6.1\pm0.2 $ dB below shot noise (and $ 15.9\pm0.2 $ dB of antisqueezing) down to 10 Hz is measured when the squeezed vacuum state is directly sent to the homodyne detector, bypassing the filter cavity. 

\subsection{\label{subsec: SUS} Filter cavity} 

The filter cavity is hosted in one of the 300-m arms of the former TAMA interferometer, and uses the already existing vacuum system.
The 10-cm diameter cavity mirrors are suspended with a double pendulum system \cite{typeC} placed on a vibration isolation multilayer stack \cite{stack}, initially developed for TAMA and recommissioned for this experiment. Mirror motion is sensed by optical levers and coil-magnet actuators are used to align the cavity and to damp suspension mechanical resonances \cite{phd_eleonora}.

In order to obtain a rotation around 30-70 Hz a storage time of about 3 ms is needed. This will require either a long cavity or a very high-finesse one. The use of order 100-m-long filter cavities has been envisaged for advanced detectors, as they are more robust to length-dependent sources of squeezing degradation such as cavity losses and residual length noise \cite {PhysRevD.90.062006,capocasa1}. The frequency at which the rotation takes place depends on the cavity linewidth, which is inversely proportional to its length and finesse. Given the 300-m length of the cavity, the finesse is chosen to be $\sim4400$ at 1064-nm wavelength, to provide a squeeze ellipse rotation at approximately 75 Hz, corresponding to an optimal quantum noise reduction for KAGRA \cite{capocasa1}. 
Since cavity losses are a major threat to squeezing \cite{PhysRevD.90.062006}, requirements on the mirror surface quality were first determined by numerical simulations, using realistic mirror maps \cite{capocasa1}. After the cavity assembly, round-trip losses were measured in the range of 50-90 ppm, compliant with requirements \cite{capocasa2}.  The filter cavity parameters are reported in Tab. \ref{recparfc}.

\noindent
\textbf{Filter cavity control - } The filter cavity is controlled by an auxiliary beam, picked off the green beam produced by the SHG. The cavity finesse for such an auxiliary beam is about 25 times smaller than that of the infrared one, to facilitate the lock acquisition process.
The green beam is superposed to the squeezed beam at an in-vacuum dichroic mirror, as shown in Fig. \ref{tamascheme}. For the initial alignment operation, a pickoff of the main laser is temporarily injected into the OPO, to copropagate with the squeezed vacuum.

The cavity is kept on resonance by acting on the main laser frequency. The error signal is obtained from the reflected green beam, using a Pound-Drever-Hall scheme, and the correction signal, processed with an analog servo, is sent to the laser PZT, with a bandwidth of about 20~kHz. An acousto-optic modulator (AOM) is placed on the green beam path prior to injection into the filter cavity. By driving the AOM at different frequencies, it is possible to change the relative frequency of the green and the squeezed beam and thus control the detuning of the latter into the cavity. The cavity is kept aligned with respect to the green beam using an error signal obtained from the transmitted power. Such a signal is generated by dithering the angular position of the cavity mirrors.

\begin{table}[!ht]
\begin{center}    
\begin{ruledtabular}
\begin{tabular}{llr}

Cavity parameter                                                                                                                        \\
\hline
Length                                                                                                                                  & $300$ m \\
Mirror diameter                                                                                                                     & $ 10 $ cm  \\
Input mirror radius of curvature                                                                                            & 438 m\\
End mirror radius of curvature                                                                                              & 445 m\\
Input mirror transmissivity  (1064 nm)                                                                                   & $0.136 \%$   \\
End mirror transmissivity    (1064 nm)                                                                                  & 3.9 ppm  \\
Finesse   (1064 nm)                                                                                                              &   $4425$\\
Input mirror transmissivity  (532 nm)                                                                                     & $0.7 \% $\\
End mirror transmissivity    (532 nm)                                                                                    &$ 2.9 \%$ \\
Finesse   (532 nm)                                                                                                               & $172$  \\
Beam diameter at waist                                                                                                       & $1.68$ cm  \\
Beam diameter at the mirrors                                                                                              & $2.01$ cm  \\

\end{tabular}
 \end{ruledtabular}
\caption{Summary of the filter cavity parameters. Mirror transmissivity and radius of curvature have been measured at the Laboratoire des Matériaux Avancés (LMA).}
 \label{recparfc}
 \end{center}
\end{table}  

\section{\label{sec:meas} Frequency-dependent squeezing measurement}

\begin{figure*}[t]
	\centering{\includegraphics[width=0.9\textwidth]{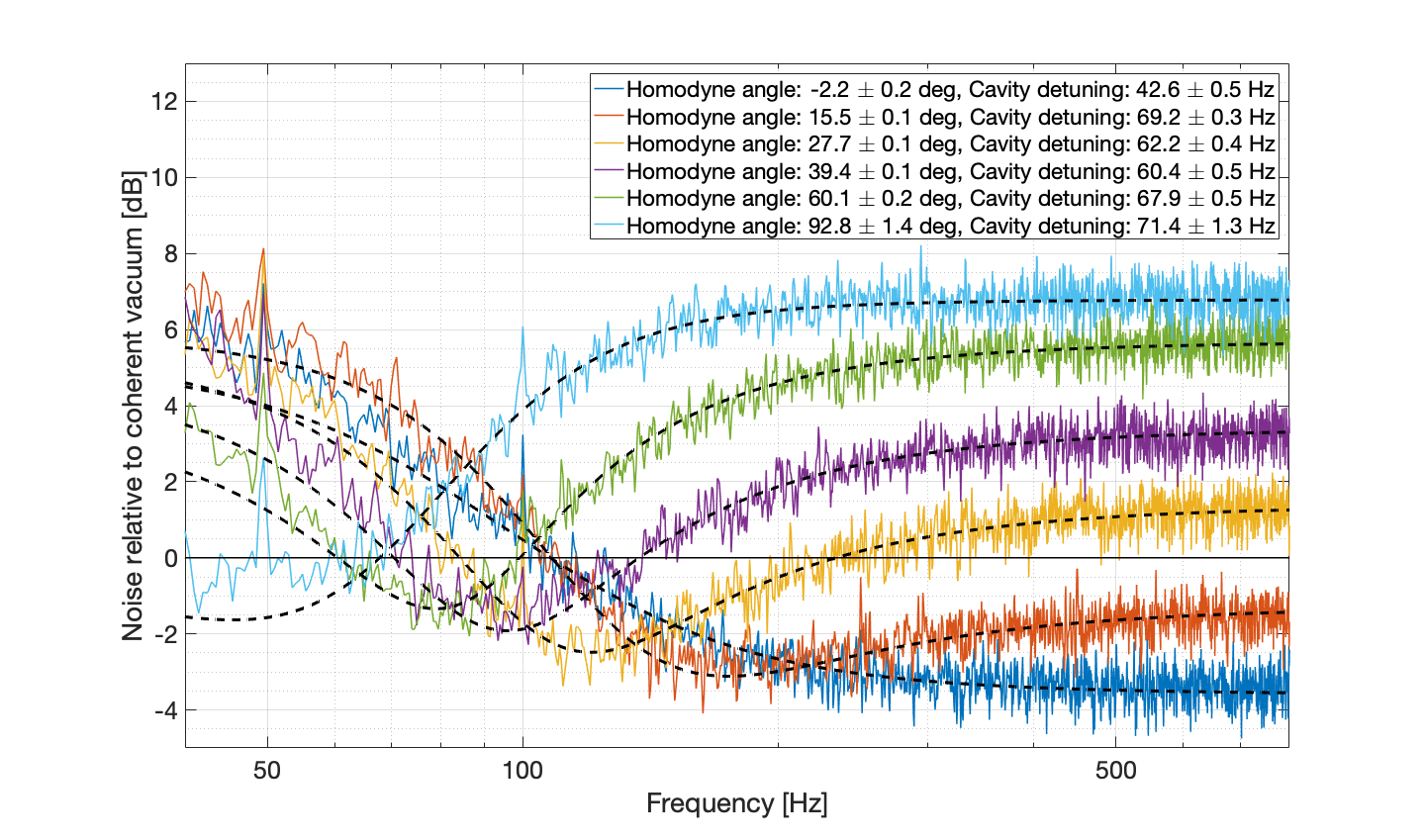}}
	\caption{Noise spectra of the frequency-dependent squeezing, measured for different homodyne angles. Each curve has been fitted, assuming the degradation parameters in Table \ref{valqui}, to extract homodyne angle and detuning frequency. The degradation parameters were used to infer the expected quantum noise reduction achievable by injecting such frequency-dependent squeezed state into a detector with the proper shot noise-radiation pressure crossover frequency. Each spectrum has a 0.5-Hz resolution and is averaged 100 times, resulting in an acquisition time of 200 s.}
	\label{FDS}
\end{figure*}

The frequency-independent squeezed beam is reflected by the cavity, acquiring a frequency dependence. This mechanism is better understood in a sideband picture, where the squeezing angle at each frequency is determined by the relative phase between the upper and lower sidebands \cite {Chelkowski}. If the cavity is operating at a detuned configuration, the symmetric sidebands experience a differential phase change depending on the cavity dispersion, resulting in a rotation of the squeeze ellipse in the quadrature plane. The relation between cavity linewidth and optimal detuning at which the cavity should be operated to obtain the desired $90 ^{\circ}$ rotation has been derived in \cite {PhysRevD.90.062006}. 

The frequency-dependent squeezed beam reflected by the cavity is sent to the balanced homodyne detector used to characterize the squeezed state. The relative phase between the squeezed beam and the homodyne local oscillator determines the direction onto which the squeezing ellipse is projected. Figure \ref {FDS} shows the measured noise spectra, normalized with respect to the nonsqueezed vacuum shot noise, for different homodyne angles.
A squeezing level of $3.4 \pm 0.4$ dB above the rotation frequency was measured, along with a squeezing level of at least 1 dB at the rotation frequency and below. In this frequency range, a lower squeezing level is expected due to two degradation mechanisms: first, the squeezing sidebands at frequencies inside the cavity bandwidth experience additional losses due to the filter cavity round-trip losses; second, an additional frequency-dependent phase noise is induced by the residual length noise of the cavity \cite{PhysRevD.90.062006}.

From the frequency dependence of the spectra it is possible to infer that the rotation of the squeeze ellipse takes place at approximately 90 Hz \cite{nota}. The curves are fitted with a model that uses the squeezing degradation parameters previously measured and reported in Table \ref{valqui}. Such parameters are the same for every curve, while the detuning and the homodyne angle are fitted for each curve.

Even if it is possible to operate the cavity at the optimal detuning, this parameter was set to have a rotation at a slightly higher frequency than the designed one (about 90 Hz instead of 75 Hz), in order to better resolve the frequency-dependent squeezing spectra in the rotation region, since the spectrum is dominated below $\sim 50$~Hz by the backscattering noise.

\begin{figure}[h!]

	\centering{\includegraphics[width=0.48\textwidth]{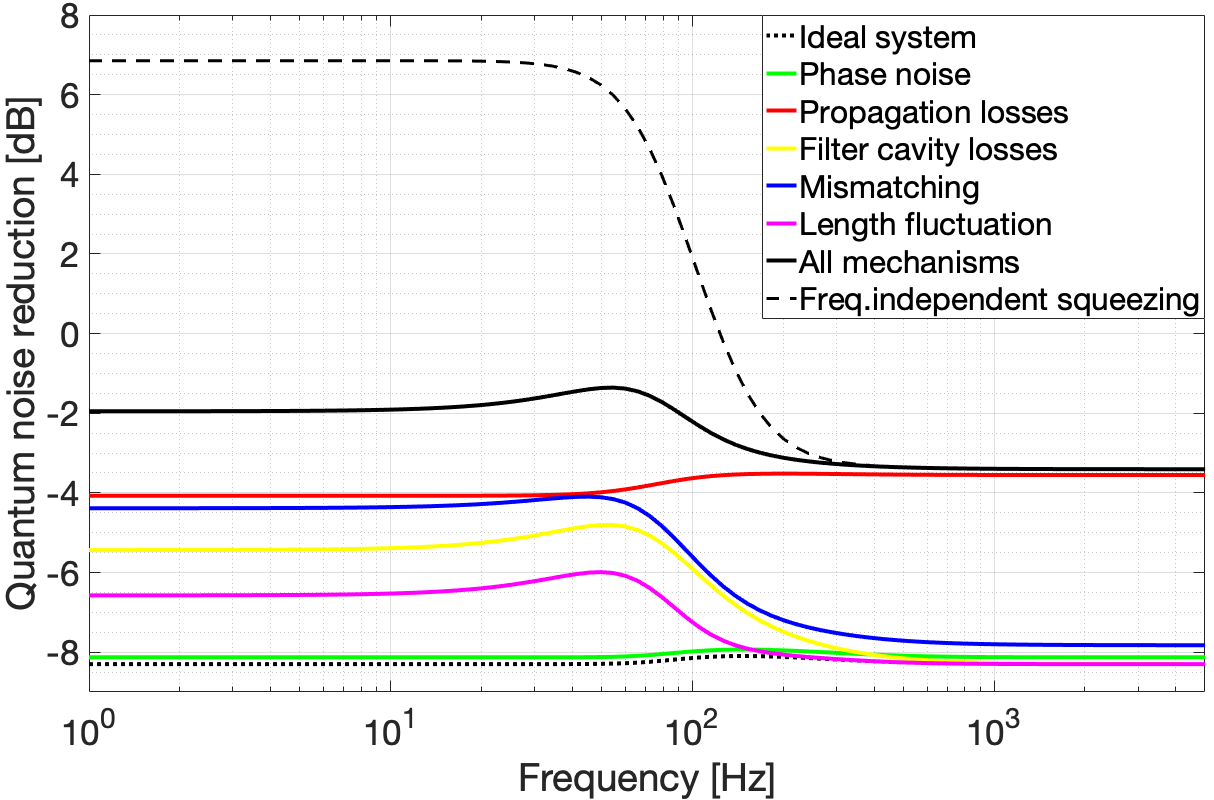}}
	\caption{Estimated degradation budget for the frequency-dependent squeezing source. The black curve shows the expected improvement of the quantum noise for a gravitational-wave detector such as KAGRA. The squeezing degradation parameters used are those reported in Table \ref {valqui}. The dashed line shows the change in the quantum noise when injecting a frequency-independent squeezing state (with the same squeezing magnitude of the frequency-dependent case). As expected, quantum noise is reduced at high frequencies, while it highly increased at low frequencies.}
	\label{SQZdeg}
\end{figure}

\noindent
\textbf{Discussion -} A degradation analysis for the measured frequency-dependent squeezing has been performed, according to the model proposed in Ref. \cite{PhysRevD.90.062006}. The squeezing degradation sources have been independently measured and used in the model to fit the curves in Fig. \ref{FDS}. The quantum noise reduction achievable with such a system is shown in Fig. \ref{SQZdeg}, assuming a squeezing angle rotation at 75 Hz, the optimal frequency for KAGRA. This curve also assumes a perfect, lossless matching with the interferometer. On the other hand, the loss values used in Fig. \ref{SQZdeg} are larger than those projected for advanced detectors in the next observation runs (including those due to a nonoptimal coupling between the squeezer and the interferometer). Therefore, the quantum noise reduction expected in advanced detectors is greater than that presented in Fig. \ref{SQZdeg} \cite{AdVirgo+, A+}.

Residual alignment fluctuations of the cavity mirrors and mirrors used to inject the beam into the cavity limit the coupling between the squeezed beam and the cavity to $94\%$. Therefore a more effective vibration isolation system would improve the performances.  

Propagation losses for the squeezed field are dominated by the low OPO escape efficiency ($ \sim 92 \%$) and by the in-vacuum Faraday isolator loss ($ \sim 14 \%$ for double pass). Both components are expected to be replaced with lower loss ones. In particular, Faraday isolators with losses below $1 \%$ \cite{Faraday} and OPO with escape efficiency as high as $99 \%$ \cite{15db} have been realized.
We expect that the integration of a more performant Faraday isolator and vibration isolation system will also reduce the backscattering  which is currently limiting the bandwidth of our measurement to $\sim 50$ Hz. 

Since the cavity is locked and kept aligned with respect to the green auxiliary beam, any relative misalignment of the squeezed beam with respect to the green beam will not be corrected. In fact, a slow drift of the squeezed beam axis with respect to the green one was observed. This limits the long-term operation of the filter cavity. We also observed a correlation between the change in the squeezed beam alignment condition and detuning. This is suspected to be the source for the detuning change observed between the squeezing spectra of Fig. \ref{FDS}. The mechanisms which couple a variation of the alignment with a detuning change are still under investigation. In order to solve these problems, an alternative strategy which uses the already present coherent control field for both length and angular control is being developed  \cite{Aritomi}. A similar solution, using an additional beam at nearly the squeezing frequency, has been successfully tested in Ref. \cite{MITpaper}, for what concerns the length control.

 \begin{table}[!ht]
 \begin{center}    
 \begin{ruledtabular}
 \begin{tabular}{llr}

Squeezing degradation parameter                                                                  & Value \\
\hline

Filter cavity losses                                                                                           & $120 \pm 30$ ppm   \\
Propagation losses                                                                                          & $36\%\pm 1\% $   \\
Mode-mismatch squeezer-filter cavity                                                             & $6\%\pm 1\% $    \\
Mode-mismatch squeezer-local oscillator                                                       & $2\%\pm 1\% $   \\
Filter cavity length noise (rms)                                                                        & $6 \pm 1$ pm \\
Phase noise                                                                                                    & $30 \pm 5$ mrad \\
Produced squeezing                                                                                       & $8.3 \pm 0.1\,\mathrm{dB}  $  \\

\end{tabular}
\end{ruledtabular}
\caption{Squeezing degradation parameters used to fit the frequency-dependent noise curves in Fig \ref{FDS}.}
\label{valqui}
\end{center}
\end{table}  

\section{\label{sec:conclusion} Conclusions} 

We have reported the first demonstration of a squeezing source able to reduce quantum noise in the whole spectrum of advanced gravitational-wave detectors.
 
The rotation of the squeezing ellipse is below 100 Hz, in the region necessary for optimal quantum noise reduction in advanced gravitational-wave detectors. This was achieved using a frequency-independent squeezed vacuum state reflected by a 300-m suspended filter cavity, operating in a detuned configuration. Filter cavities of comparable length are expected to be integrated soon in KAGRA, Advanced LIGO, and Advanced Virgo. As such, the results presented here constitute a key validation for the use of this technology in advanced detectors.

Current limitations to the measured squeezing are well understood, as are the solutions to mitigate them, which should allow to obtain at least 4 dB of broadband quantum noise reduction \cite{capocasa1}. However, the frequency-dependent squeezing source described in this Letter, at the present stage of development, would already allow for a broadband reduction of quantum noise in a detector like KAGRA. Since quantum noise is dominating the design sensitivity above 30 Hz and is anyway comparable with other noise sources below that frequency, the use of frequency-dependent squeezing would result in an overall sensitivity improvement \cite{capocasa1}. 

This result is not only a step toward the implementation of planned upgrades to current gravitational-wave detectors, but also a demonstration of planned technology for use in third-generation detectors such as Einstein Telescope, which features frequency-dependent squeezing with long filter cavities \cite{ET}.

\FloatBarrier

\section*{Acknowledgements}
We thank R. Schnabel, D. Tatsumi, E. Schreiber, L. Pinard, K. Somiya, J. Degallaix, S.R. Wu, Y. Enomoto, L. Trozzo, S. Zeidler, M. Marchiò, N. Hirata, I. Fiori, P. Ruggi, F. Paoletti, C. De Rossi, T. Akutsu, T. Tomaru, E. Majorana, K. Izumi, M. Mantovani, and J. Baird for the useful contributions and discussions. We thank S. Oshino, T. Yamamoto, and Y. Fujii for the help with the digital control system. We thank also the Advanced Technology Center (ATC) of NAOJ for the support. This work was supported by the JSPS Grant-in-Aid for Scientific Research (Grants No. 15H02095 and No. 18H01235), the JSPS Core-to-Core Program, and the EU Horizon 2020 Research and Innovation Programme under the Marie Sklodowska-Curie Grant Agreement No. 734303. N.A. was supported by JSPS Grant-in-Aid for Scientific Research (Grants No. 18H01224 and No. 18K18763) and JST CREST (Grant No. JPMJCR1873).


\begin{thebibliography}{99}
\bibitem{nota}
The 90 Hz rotation value corresponds to a filter cavity detuning of about 63 Hz, which is the average of the fitted detuning values of all the spectra shown in Fig. \ref{FDS}.
\end{thebibliography}
\end{document}